\RequirePackage[2020-02-02]{latexrelease}
\documentclass[twocolumn,prl,amsmath,amssymb,showpacs,superscriptaddress,floatfix]{revtex4-2}
\usepackage{graphicx}
\usepackage{bm}
\usepackage{lineno}
\usepackage{epsf}
\usepackage{xcolor}
\begin{document}
\title{Kob-Andersen model crystal structure: genetic algorithms vs spontaneous crystallization}

\author{Yu. D. Fomin \footnote{Corresponding author: fomin314@mail.ru}}
\affiliation{Vereshchagin Institute of High Pressure Physics,
Russian Academy of Sciences, Kaluzhskoe shosse, 14, Troitsk,
Moscow, 108840, Russia }

\author{N. M. Chtchelkatchev}
\affiliation{Vereshchagin Institute of High Pressure Physics,
Russian Academy of Sciences, Kaluzhskoe shosse, 14, Troitsk,
Moscow, 108840, Russia }

\date{\today}

\begin{abstract}
For the first time, the crystal structure of the Kob-Andersen mixture has been probed by genetic algorithms calculations. The stable structures of the system
with different molar fractions of the components have been identified and their stability at finite temperature has been verified. A possibility
to obtain these structures by spontaneous crystallization of a liquid has been checked.

\end{abstract}


\maketitle


\section{Introduction}

There is an assumption circulating in the scientific community that regions with good glass-forming ability in the phase diagram correlate with regions where many metastable states lie near the convex-hull (in other words, the convex hull is not separated by a ``gap'' from metastable states). We considered the Kob-Andersen model to test this conjecture. Using evolutionary algorithm we built the convex hull for this model and found the energy distribution of metastable states. 

The Kob-Andersen (KA) model \cite{Kob1994,Kob1995,Kob1995a,Coslovich2024} is probably the most well recognized theoretical model used to study the glass transition. Due to its simplicity it appeared to be a very convenient model to study the viscous slowing down of supercooled liquids both theoretically (for instance, within a Mode Coupling
Theory formalism \cite{Kob1995,Kob1995a}) and in molecular dynamics simulation \cite{Kob1994}.

The KA mixture is a mixture of particles interacting via the Lennard-Jones (LJ) potential:
\begin{equation}
  U_{LJ}=4 \varepsilon_{\alpha \beta} \left( (\sigma_{\alpha \beta}/r)^{12}-(\sigma_{\alpha \beta}/r)^6 \right),
\end{equation}
where $\alpha$ and $\beta$ are $A$ or $B$ for the different LJ species of the mixture. The porential is truncated and shifted at $r_c=2.5 \sigma_{AA}$. The potential parameters
are taken as $\varepsilon{AB}/\varepsilon_{AA}=1.5$, $\varepsilon{BB}/\varepsilon_{AA}=0.5$, $\sigma_{BB}/\sigma_{AA}=0.88$ and $\sigma_{AB}/\sigma_{AA}=0.8$. Most of the
studies consider the KA mixture with $80 \%$ o thef A particles and $20 \%$ of the B ones at the reduced density $1.2$ in the units of particles A.

Usually when the glass transition is studied it is somehow related to the stable crystal phase in the system. For instance, the degree of supercooling of the liquid, the time required to
nucleation for the glass, etc. are discussed. However, finding the crystal structure of the KA model appeared to be a very difficult task. Crystallization behavior of the binary LJ mixture was
studied in Ref. \cite{Th2001} and it was shown that the cubic close packed crystal has lower energy than a disordered state. However, the ground state crystal structure has not been
recognized in this paper.


Another attack on the problem was undertaken in Ref. \cite{Fernandez2003}. The authors of this paper guessed several crystal structures and compared their stability
by performing a conjugate gradient minimization of energy (at constant volume) or enthalpy (at constant pressure). They found that equimolar mixture easily crystallizes into a
$CsCl$ structure. At some concentrations the system can split into a mixture of a $CsCl$ and an FCC crystals. However, the system with $80 \%$ of the A particles fails to crystallize.

In Ref. \cite{Toxvaerd2009} the KA system was modified to simplify its crystallization. Based on these results the authors concluded that the KA mixture should nucleate after 0.1 ms (in
units of Argon). Therefore, long simulations are required to observe the crystalization of the KA mixture. They also managed to find the FCC and the $CsCl$ structures in the mixtures with
the fraction of the B component.

In Ref. \cite{Pedersen2018} a molecular dynamics study of phase diagram of a binary LJ mixture with the interaction parameters of the KA mixture was performed.
The authors employed a two-phase simulation method, i.e., crystal and liquid coexisted in the simulation box and melting or crystallization of the system
was monitored to find the limits of stability of the phases. A half of the box consisted of a KA liquid, while another half was occupied by a some crystal phase.
Exchanging of the B particles between the subvolumes allowed to change the concentration in both parts of the system. The authors identified several possible crystal structures
in the systems with different molar fractions of the components, line, FCC, $PuBr_3$ structure, $Fe_3C$, $Ni_3P$, $Al_2C_4$ and $CsCl$ ones.

The structure of the KA nanoparticles was explored in Ref. \cite{Zh.NChen2018}. Although the structure of nanoparticles can be strongly different from the one of the bulk system
the authors also find that the system forms an FCC crystal at low molar fraction of the B species and a $CsCl$ crystal when the molar fraction of the B particles aproachs $0.5$.

It is seen from the discussion above that all previous studies (to the best of our knowledge) of crystal structure of the KA mixture are based either on spontaneous crystallization of the system or on
consideration of some of anticipated structures. However, it is well known that these methods are not enough efficient for finding the ground state of the system \cite{Oganov2011}.
The most powerful method of finding of stable crystal structures is the one based on genetic algorithms \cite{Oganov2011}. Application of genetic algorithms allowed to
obtain numerous unexpected results, including formation of compounds with unusual composition, for instance, a $NaCl_3$ \cite{Zhang2013,Saleh2016}, prediction of structure of
nanoparticles \cite{Lyakhov2013}, structure of surfaces \cite{Wang2014}, etc.

The goal of the present paper is to explore the stable crystal structure of a binary LJ mixture with interaction parameters of the KA model in the whole range of
concentrations by means of the genetic algorithms. We also perform a molecular dynamics simulation of spontaneous crystallization of the system in order to
see whether the structures obtained within the framework of genetic algorithms can be obtained by spontaneous crystallization of a liquid.

\section{System and Methods}

\subsection{Genetic algorithms calculations}
The search for stable structures was carried out using the genetic evolutionary algorithm USPEX~\cite{Oganov2011,Lyakhov2013} over the entire concentration range. 
The maximum number of particles in a cell was 80. Energy minimization was carried out using the LAMMPS package. 
During the evolutionary search, about 100000 different crystal configurations were considered.

\subsection{Molecular dynamics simulations}

Two sets of molecular dynamics simulations was performed. In the first set we simulated the crystalline structures obtained by genetic algorithms calculations
at finite temperature $T=50$ K to check their stability. 

Since the genetic algorithms calculations are performed in physical units (the units of energy are eV, the units of length are $\AA$, etc.), we employed the
same physical units in molecular dynamics simulations (units 'metal' in lammps). The LJ parameters for particles A are given in argon units: $\varepsilon_A=119.8$ K,
$\sigma_A=3.504$ $\AA$. All other interaction parameters were obtained according to the KA mixture rules. Mass of all particles in the system was taken as
mass of argon atoms: $m_A=m_B=39.948$ a.m.u. 

The number of particles depended on the structure and varied between 3456 and 8000. Periodic boundary
conditions were imposed. The Nose-Hoover thermostat was employed to fix the temperature. The timestep was set to $1.0$ fs and the total simulation time was
$1.0$ ns, which should be sufficient for the structure collapse if it is unstable.

In the second set of simulations we simulated a possibility of a liquid mixture with different concentration of components to experience spontaneous crystallization.
A system of 32000 particles in a cubix box with periodic boundary conditions was used. Firstly we simulated the systems from pure component A to pure component B with the
step in molar fraction of B $\Delta c_B=0.1$. Later on we simulated the concentrations from $c_B=0.5$ to $0.8$ with smaller step $\Delta c_B=0.02$.
The system was simulated at constrant pressure $P=425$ bar, which corresponds to the reduced pressure $P*=10.19$ which was used in Ref. \cite{Pedersen2018}.
The initial structure was an FCC solid with random distribution of the A and B species across the lattice cites. The crystal was melted at high temperature $T_h=500$ K.
All systems were liquid at this temperature. Later on the system was immediately quenched to lower temperature $T=100$ and $T=50$ K. The system was
simulated at constant pressure and constant temperature for $1.0$ ns to find out the equilibrium density. The structure of the system is also monitored to see
whether it experience a spontaneoud crystallization. After that the system was deformed to reach the equilibrium density and simulated for $1.0$ ns
in a NVT (constant number of particles N, volume V and temperature T) ensembe and more $1.0$ ns in a NVE (constant internal energy E) ensemble. During
these simulations we also controll whether the system is in a liquid state liquid and calculate its structural properties, such as partial radial distribution functions (RDFs),
the bond orientational orders (BOO) \cite{Steinhardt1983}, the number of nearest neighbors. The number of nearest neighbors was calculated via a Voronoi construction.
We also characterize the dynamic properties of the system via its shear viscosity by the Green-Kubo method.

Molecular dynamics simulations were performed using the LAMMPS simulation package
\cite{Plimpton1995}.

\section{Results and Discussion}

\subsection{An approach from solid}

First of all we make a search of stable crystal structures of the KA mixture in the whole range of concentrations of the components. The convex hull of
the system is shown in Fig. 1. It is seen that the stable crystal structures are formed at six different concentrations. These results are summarized in
Table I. As it follows from Table I, the stable crystalline structures appear when the molar fraction of the components is a ratio of integers: $0$, $1/2$, $2/3$, $3/4$ and $1$.
The only exception is $c_B=0.58$. However, closer observation of this structure shows that it is a defected crystal, i.e., although it has a negative formation
energy it should be considered as a low energy metastable crystlal (see below). Below we discuss the structures at all mentioned concentrations in details. The stable
crystal at zero and unity concentration $c_B$ is the FCC one. This result looks trivial and we do not discuss it below.

\begin{figure}[htb]
\centering
\includegraphics[width=8cm]{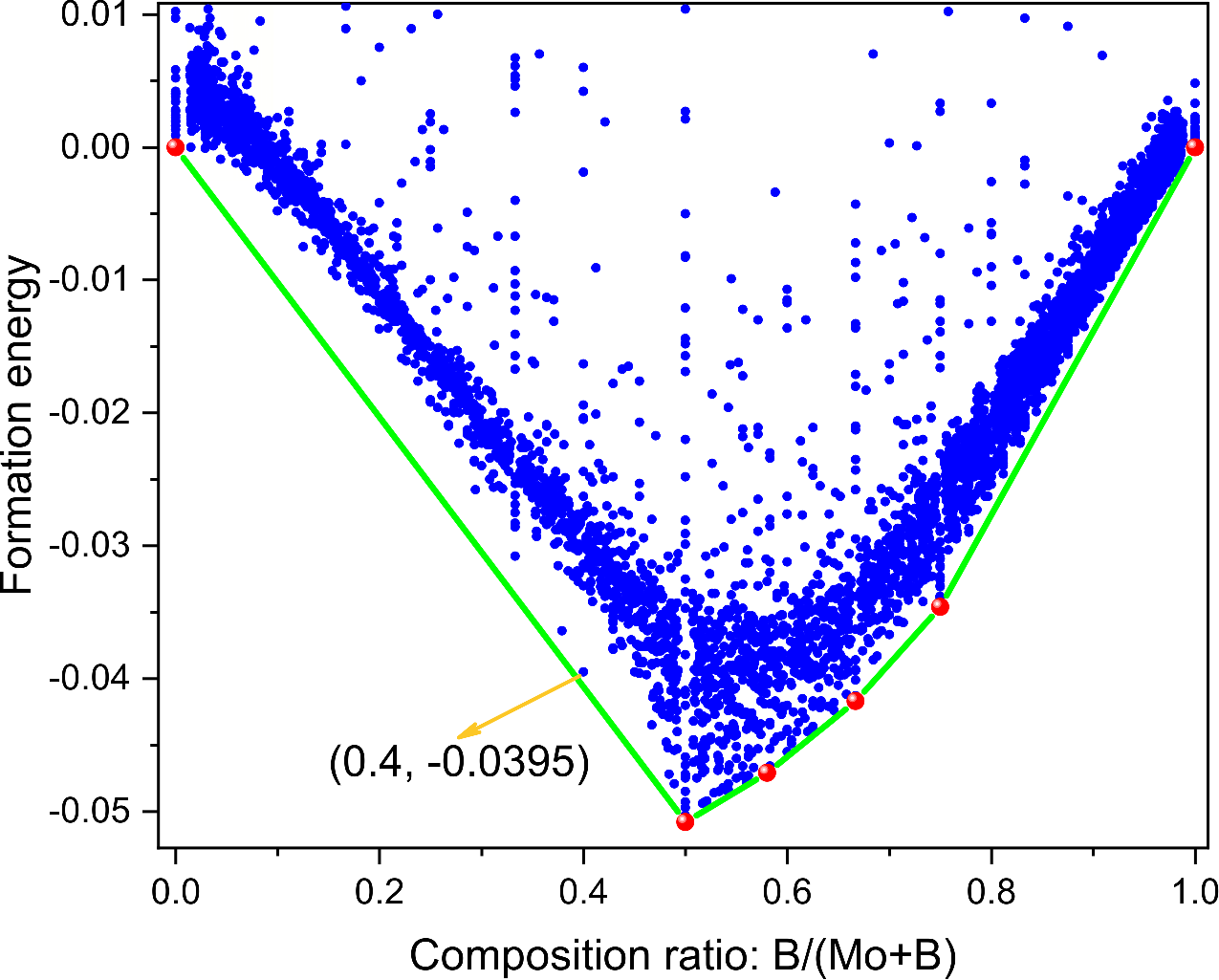}%
\caption{\label{ch} A convex hull construction of the KA mixture. The red balls show stable structures; we discuss them in the text. The insert highlights a metastable structure close to the convex-hull in a region where there are no stable structures in a wide concentration range. }
\end{figure}

\begin{table}
\begin{tabular}{ |c| c| c| c|}
\hline
 $c_B$ & Symmetry Group & V ($\AA^3$)  & E (eV)\\
\hline
 0.0 & 225 & 40.0216 & 0.0 \\
\hline
 0.5 & 221 & 23.9961 & -0.0508 \\
\hline
 0.58 & 1 & 23.5558 & -0.0471  \\
\hline
 0.667 & 1 & 23.1472 & -0.0417 \\
\hline
 0.75 & 194 & 23.011 & -0.0346 \\
\hline
 1.0 & 225 & 27.2116 & 0.0 \\
\hline
\end{tabular}
\caption{The stable crystal structures of the KA mixture at different molar fraction of component B $c_B$. The third column gives volume per atom and the last column gives the formation energy.}
\end{table}


The crystal structure of the mixture at equal concentration of the components is shown in Fig. 2. It is seen that the system forms a cesium chloride crystal structure,
which is consistent with previous publications of different authors \cite{Fernandez2003,Pedersen2018}.
We perform a molecular dynamics simulation at $T=50$ K in order to verify its stability at finite temperature and find that the structure preserves after $1.0$ ns. We believe
that this run is sufficient to confirm that the structure is stable.

\begin{figure}[htb]
\centering
\includegraphics[width=8cm]{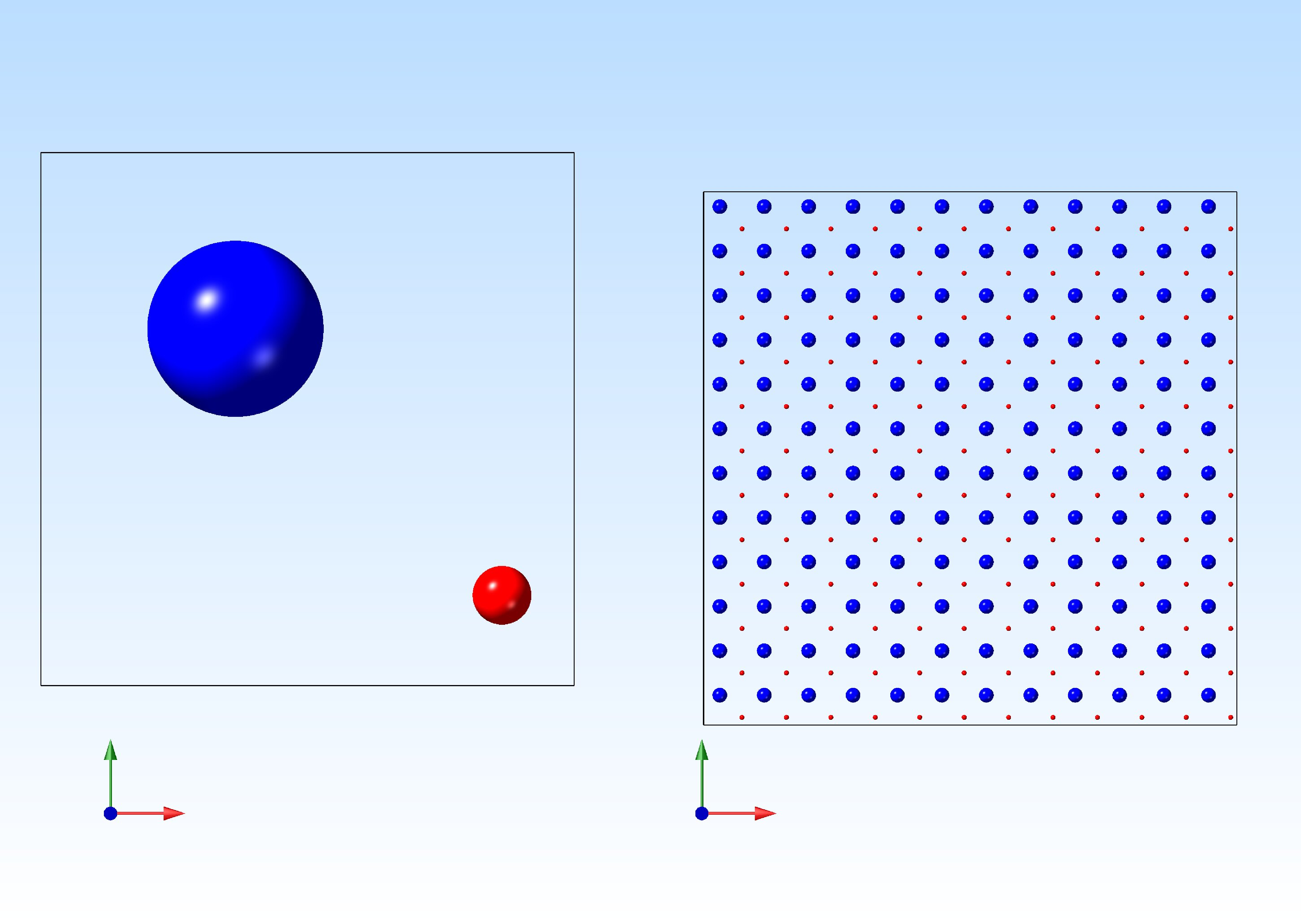}%
\caption{\label{c05} Crystal structure at the concentration $c_B=0.5$. Big blue balls show type A particles while small red ones - type B.}
\end{figure}

Predicted crystal structure at $c_B=0.58$ is shown in Fig. 3. It is clearly seen from the right part of this figure, that the lines of both A and B particles are curved,
which states that this structure does not indeed poses translational invariance. As a result we conclude that this structure is a low energy metastable phase. At the same time
we find that this structure is stable in molecular dynamics simulation at $T=50$ K. Most probably, this structure can be refined to find a stable crystal which, however, requires
a lot of computational resourses.

\begin{figure}

\includegraphics[width=8cm, height=6cm]{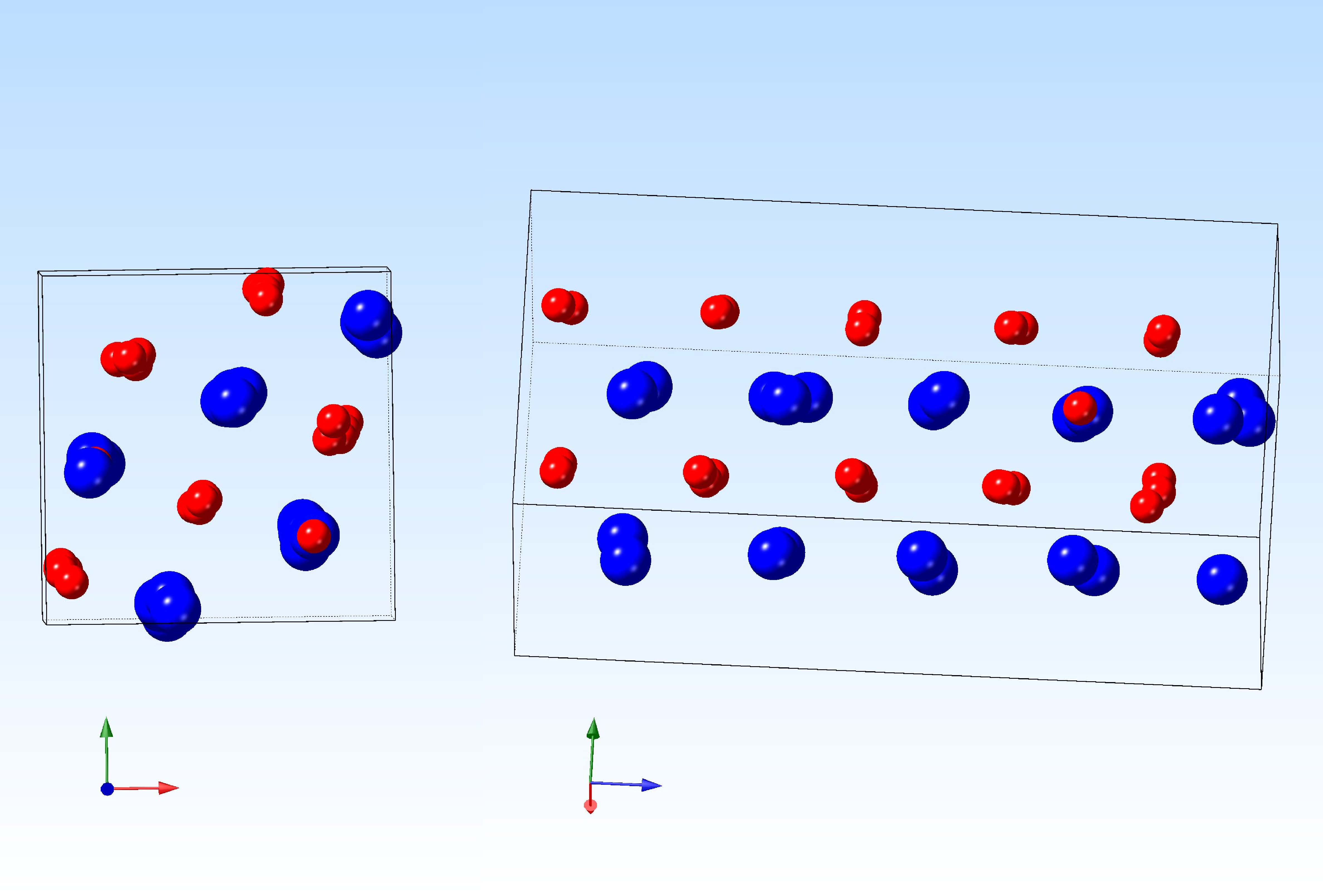}%

\caption{\label{c058} The crystal structure at the concentration $c_B=0.58$ from two different angles. Big blue balls show type A particles while small red ones - type B.}
\end{figure}

The structure of the system at $c_B=2/3$ is given in Fig. 4. This structure is also stable in molecular dynamics simulation at $T=50$ K.

\begin{figure}[htb]
\centering
\includegraphics[width=8cm]{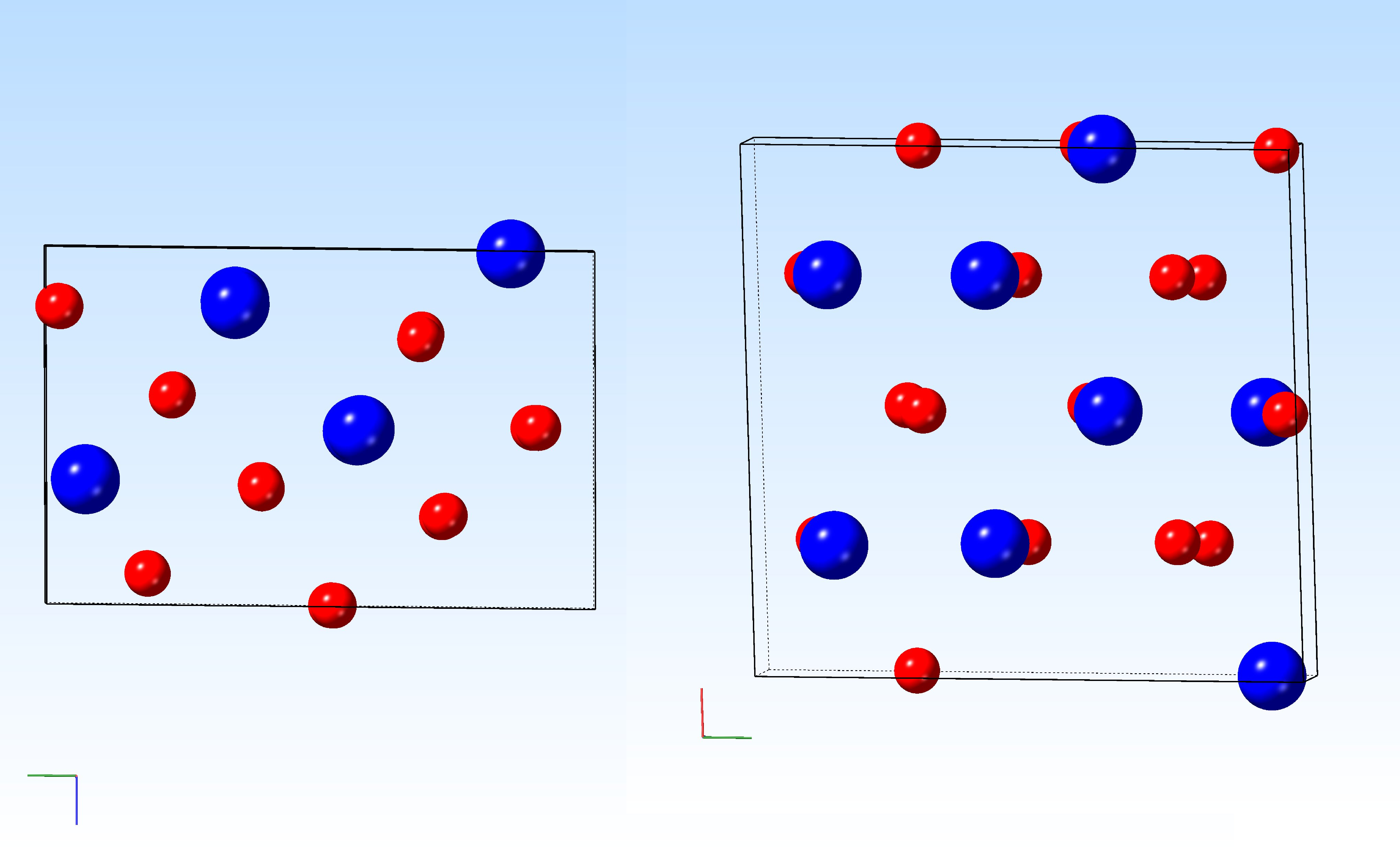}%
\caption{\label{c0667} The crystal structure at the concentration $c_B=2/3$. Big blue balls show type A particles while small red ones - type B.}
\end{figure}

Finally, we show the structure of the system at $c_B=3/4$ (Fig. 5). This structure becomes inclined: the angles $\alpha=\beta=90^o$, $\gamma=120^o$. It can be represented
as two sublattices of hexagons: the hexagons of the B particles are located inside the hexagons of the A ones. It is also important to note, that the volume per particle decreases with concentration up to
$c_B=0.75$, but increases to $c_B=1.0$. Therefore, a minimum of the volume per particle takes place at this concentration.

\begin{figure}[htb]
\centering
\includegraphics[width=8cm]{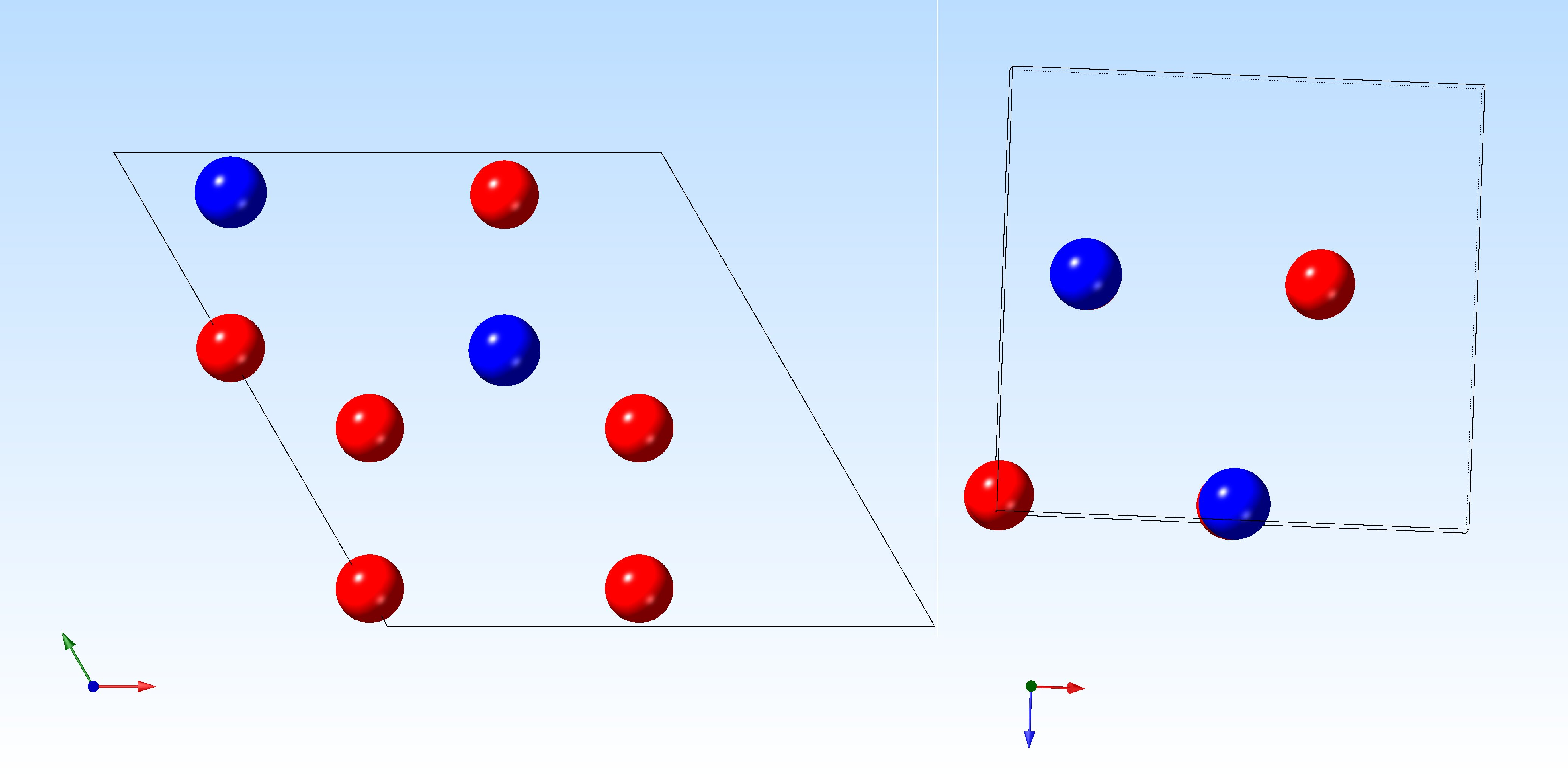}%
\caption{\label{c075} The crystal structure at the concentration $c_B=3/4$. Big blue balls show type A particles while small red ones - type B.}
\end{figure}

As one can see from the convex hull (Fig. 1), there are also many other structures with low formation energy, which, however, are not at the convex hull. We checked some
of these structures in molecular dynamics simulation at $T=50$ K and found that many of them are stable in simulation. Therefore, the system demonstrates a large number
of metastable structures with low energy, which makes it a good glassformer.

We give the POSCAR (VASP format) files with the structures under discussion in Supplementary Materials.

From Fig. 1 one can see that there is a structure at $c_B=0.4$ which is very close to the convex hull, but  does not reach it.  One can assume that this
is a metastable structure with very low formation energy. We show this structure in Fig. 6. It contains 20 atoms in a unit cell (12 A atoms and 8 B ones). 
A structure with a such large unit cell looks to be unprobable. However, it is stable in molecular dynamics simulation at
$T=50$ K.

\begin{figure}[htb]
\centering
\includegraphics[width=8cm]{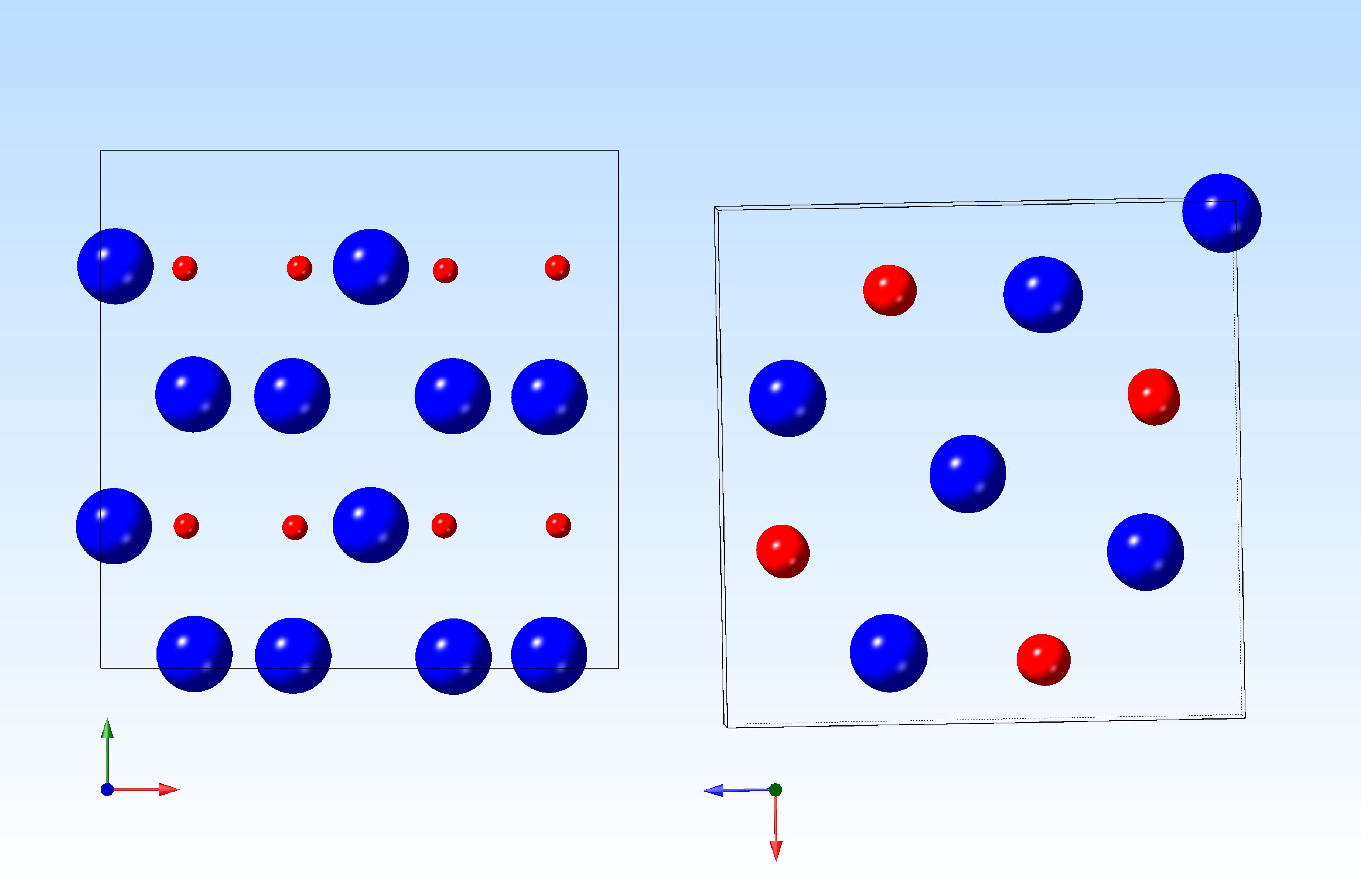}%
\caption{\label{c04} The crystal structure at the concentration $c_B=0.4$. Big blue balls show type A particles while small red ones - type B.}
\end{figure}

It is seen that several different crystalline structures were obtained in the system. These structures can be analized
by means of Smooth Overlap of Atomic Positions (SOAP) descriptors, which characterize the local environment of atoms
using the decomposition of the density of atoms according to an orthonormal basis, which includes
Gaussians, spherical harmonics and radial basis functions \cite{soap1,soap2,soap3,soap4}. These descriptors are local, i.e. they
characterize each atom of the system. However, it is possible to average these descriptors by cell and
obtain a global descriptor characterizing the entire system of atoms. This procedure was carried out. In
the Fig. 7, each point corresponds to the structure found by the genetic algorithm. The color determines
how far the structure is from the convex hull in terms of energy. The bright red color corresponds to the
convex hull. Blue dots are the most metastable structures. Since the SOAP descriptor vector contains
several dozen components, the principal component analysis (PCA) was used to compare different
structures. It can be seen that the structures are clearly divided into two "families". One cloud grows
from $c_B=0$, the other include structures $0.4 \leq c_B \leq 1.0$ . This means that, from the perspective of the local
environment, the crystal structures of the same manyfold are similar (one can switch from one structure
to another in a sequence of small changes to the local structure).

\begin{figure}[htb]
\centering
\includegraphics[width=10cm]{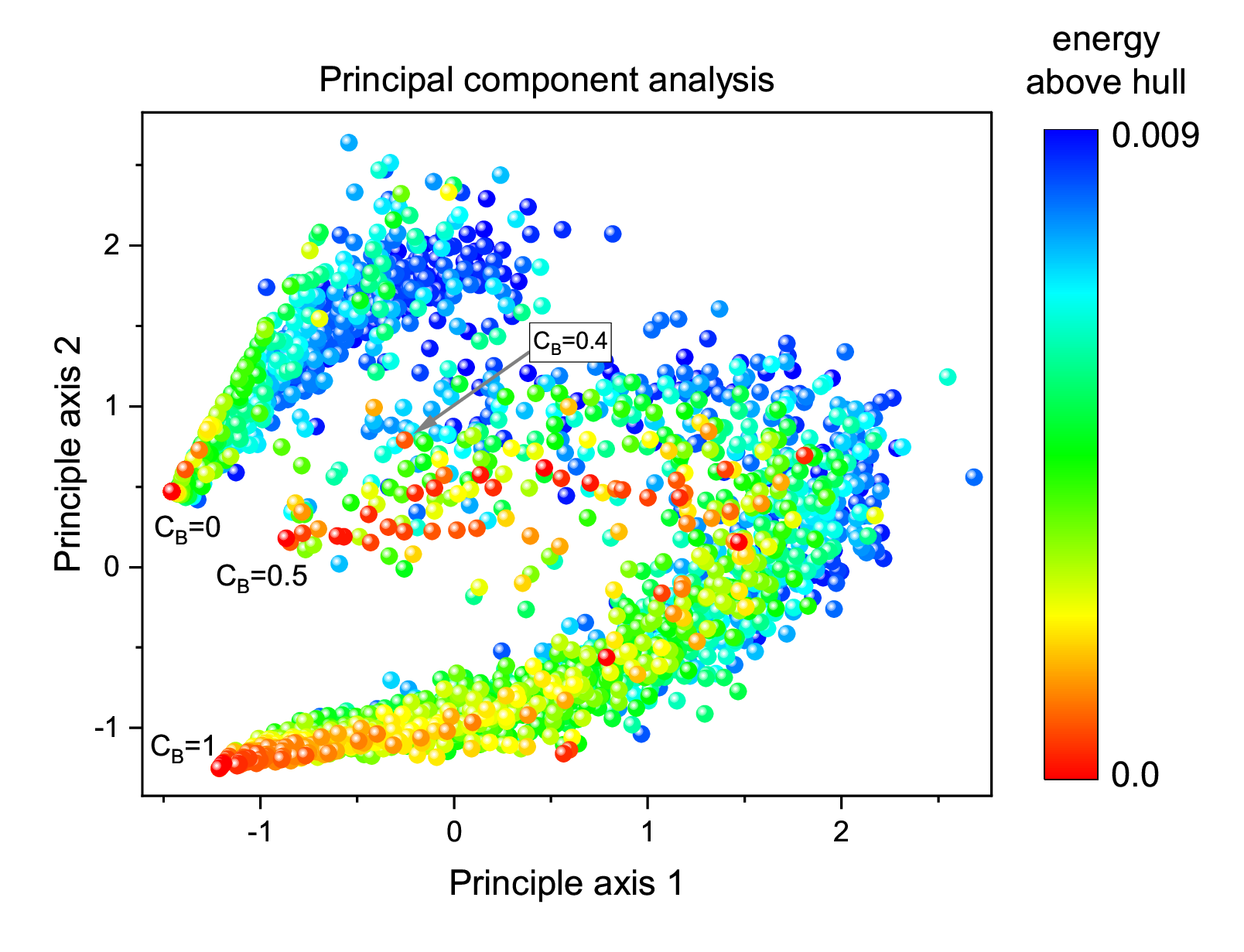}%
\caption{\label{soap} Principle components analisys of crystal structures obtained by genetic algorithms.}
\end{figure}

Importantly, evolutionarry algorithm does not find any stable crystal structure at the molar fraction $c_B=0.2$, which is used for simulation of glass forming systems (in most
of the papers on the KA mixture the molar fraction is expressed as $c_A=0.8$. We used $c_B$ instead). Moreover, all structures at this composition
are far from the convex hull, i.e. they are strongly metastable. 
Taking into acount the results of SOAP descriptors calculations, we conclude that most probably
the thermodynamically stable phase of KA system should be a mixture of FCC and CsCl structures, which is consistent with previous studies \cite{Toxvaerd2009,Pedersen2018,Zh.NChen2018}.


As it was mentioned in Introduction, it is usually assumed that a system be easily glassified if there are numerous metastable structures with similar formation energies:
the competition between these structures prevents formation of macroscopic crystals. As it is seen from the genetic algorithms calculations there is
no any crystal structure close to the convex hull at the concentration $c_B=0.2$ which corresponds to the KA mixture. At  the same time this system demonstrates 
good glass forming ability.  Therefore, this assumption is not a necessary condition for good glass forming ability of a liquid.

\subsection{An approach from liquid}

In the last part we simulate the mixture of the LJ particles at different concentrations of components starting from high temperature liquid and cooling it down.
On one hand such a method can result in spontaneous crystallization of the system. Otherwise it should lead to glass formation of the mixture. We intend
to check whether the system can crystallize into the structures discussed above.


Figure 8 shows a density of the systems in the liquid state at $T=100$ K and $P=425$ bar at different concentrations of components. It is seen that density is not
monotonous function: it increases up to $c_B=0.6$, passes a maximum at this value of the concentration and decreases at higher $c_B$.

\begin{figure}[htb]
\centering
\includegraphics[width=8cm]{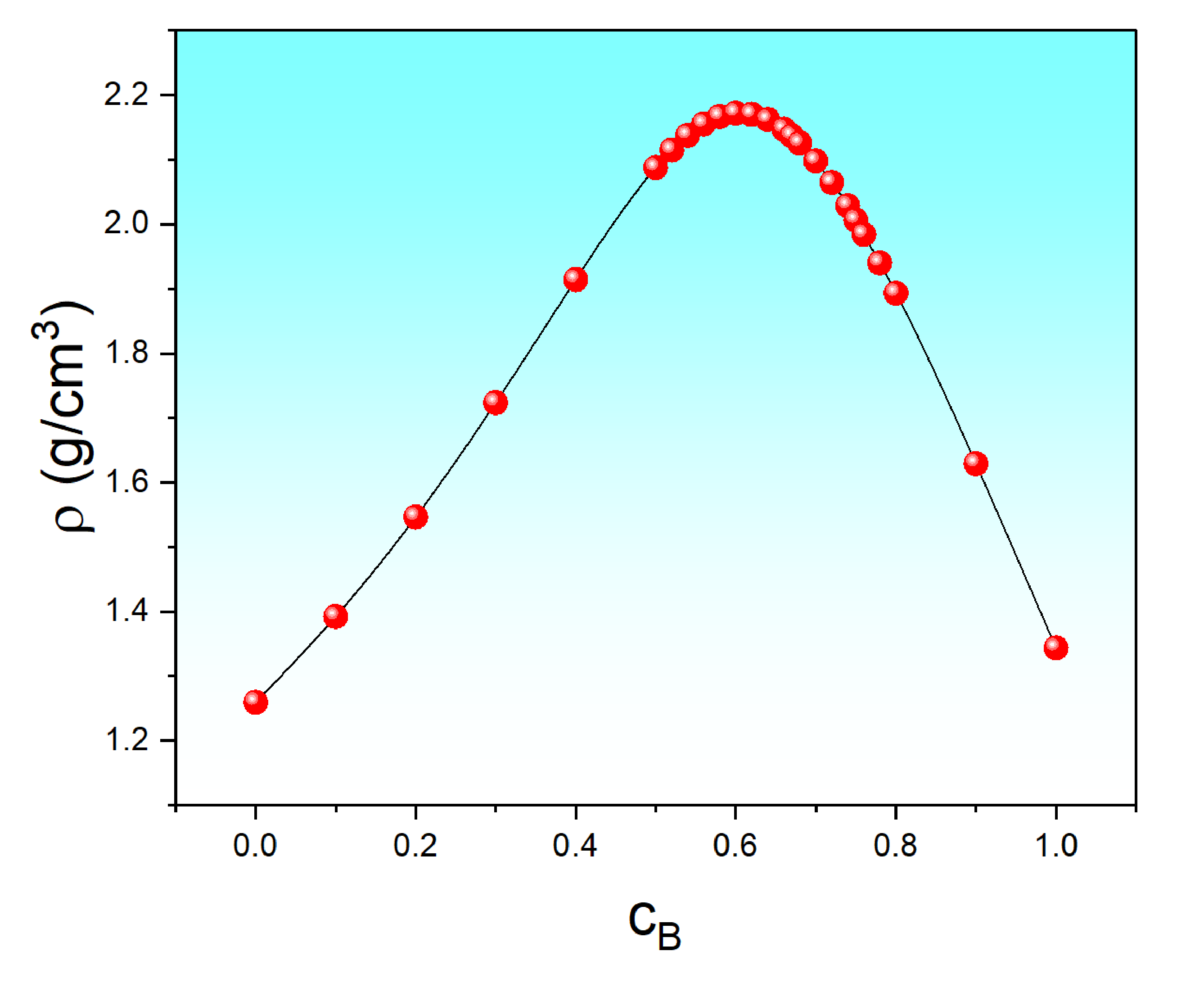}%
\caption{\label{t100} Density as a function of the B component molar fraction at $T=100$ K and $P=425$ bar.}
\end{figure}

\begin{figure}[htb]
\centering
\includegraphics[width=8cm]{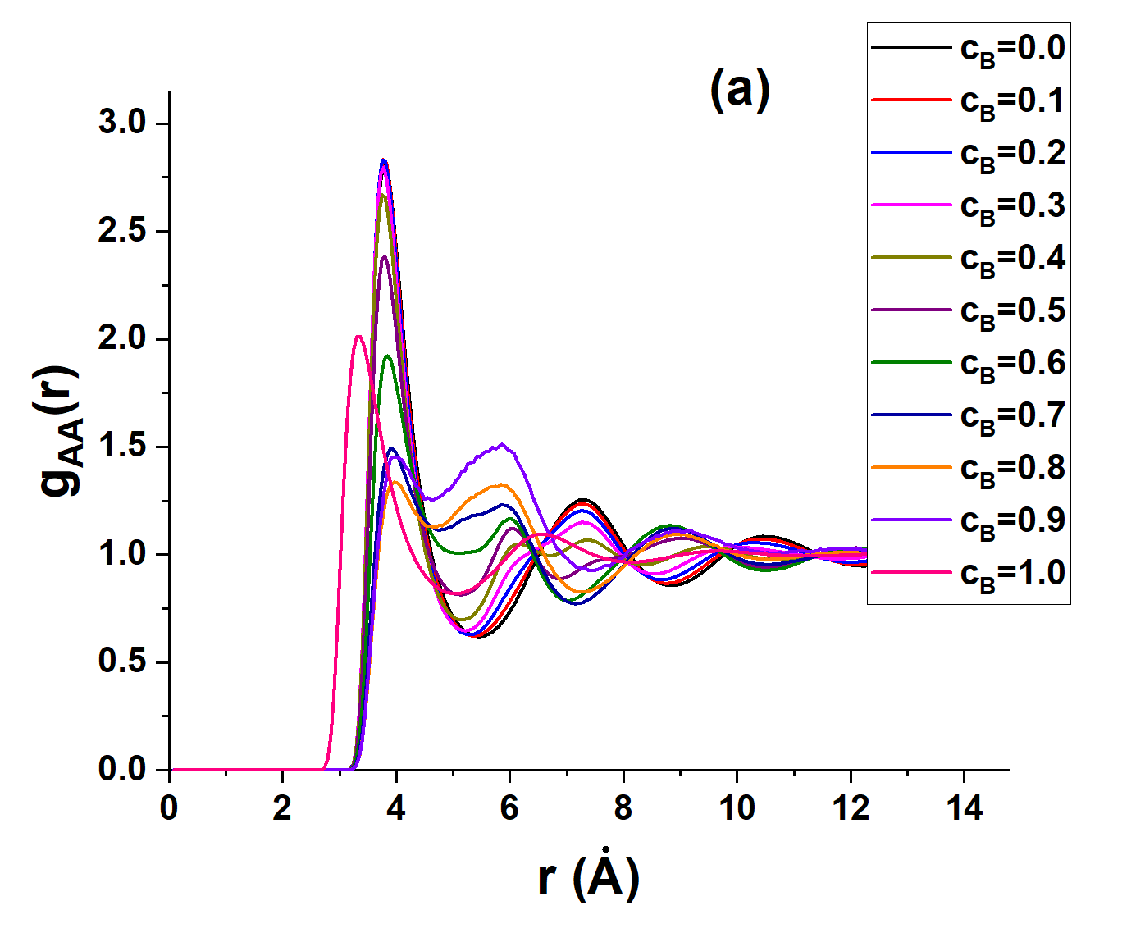}%

\includegraphics[width=8cm]{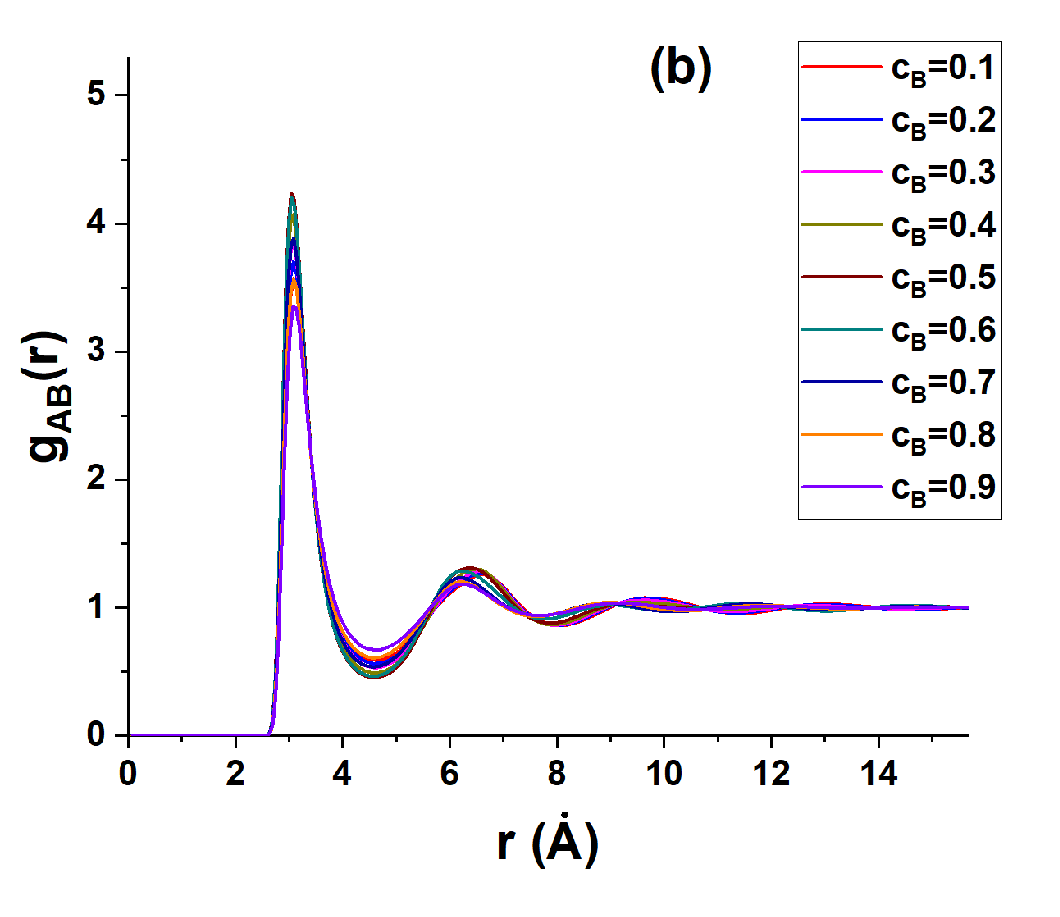}%

\includegraphics[width=8cm]{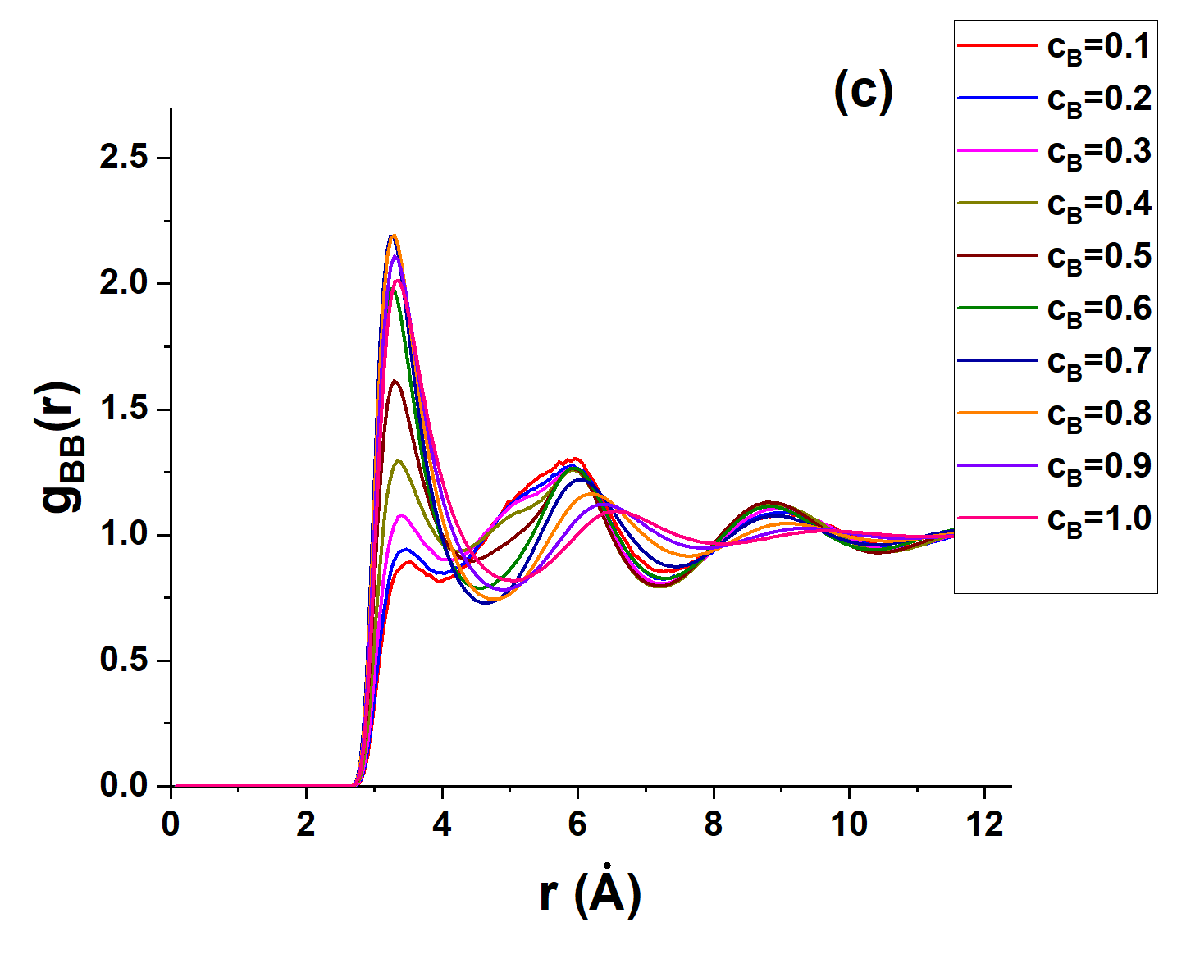}%

\caption{\label{g-100} Partial radial distribution functions at $T=100$ K and $P=425$ bar at different concentrations.}
\end{figure}

\begin{figure}[htb]
\centering
\includegraphics[width=8cm]{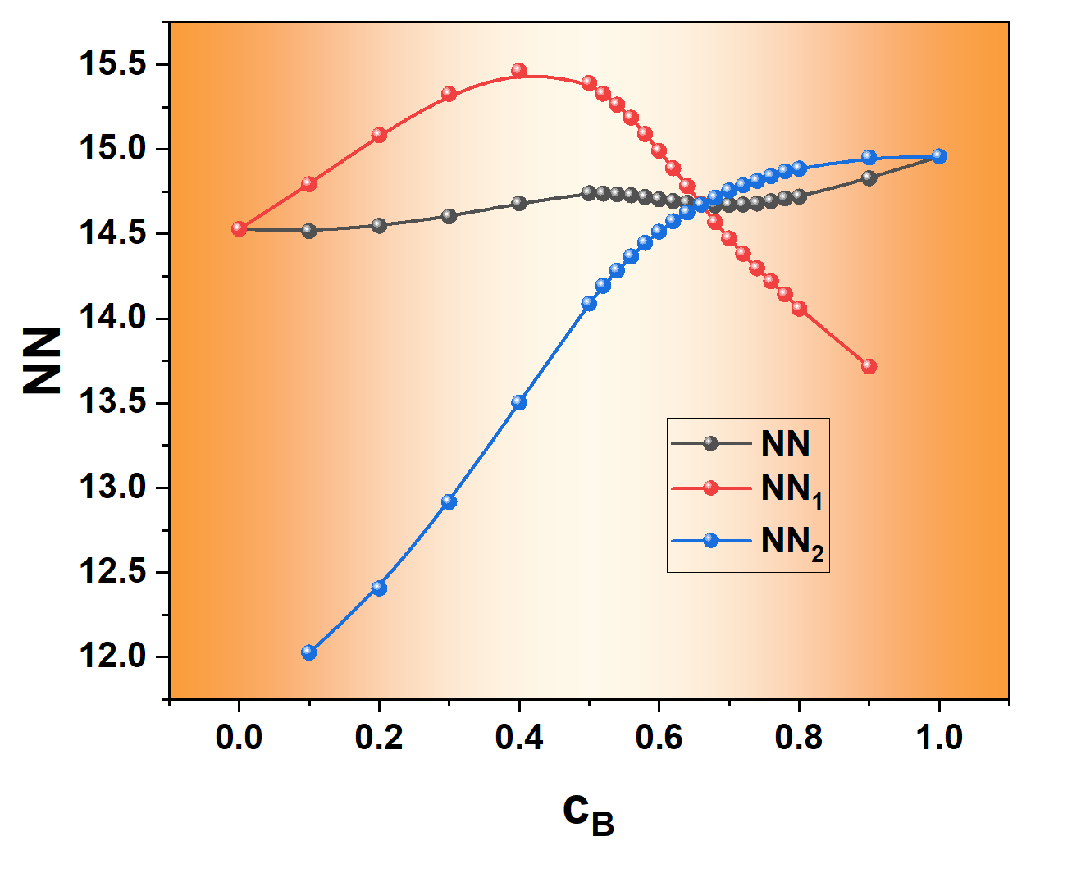}%

\caption{\label{nn} Number of nearest neighbors of the atoms at $T=100$ K and $P=425$ bar as a function of molar fraction $c_B$. NN - average number of nearest neighbors, NN1 - number of
nearest neighbors of A particles and NN2 - number of nearest neighbors of B ones.}
\end{figure}

Figure 9 shows partial RDFs of the system at $T=100$ K and different concentrations of the species. All of these RDFs
show liquid-like behavior. However, for more elaborated evaluation of the structural properties we calculate also the BOOs of the system.
We also calculate the same BOO for all crystalline structures described above and the HCP structure, which is very close in free energy
to FCC and, as a result, when an LJ liquid spontaneously crystallizes, it forms a mixture of the FCC and the HCP crystals. The same procedure is
undertaken also at $T=50$ K to monitor the effect of the temperature on the structure of the system. Below we show the results for $T=50$ K
because the crystalline systems were simulated at this temperature.



In table II we give the values of parameters $q_4$, $q_6$, $q_8$, $q_{10}$ and $q_{12}$ for all solid
structures observed in the genetic algorithm calculations above and the HCP crystal. We calculated the same
set of the BOO parameters in simulations of liquid mixtures. Each particle in the liquid is considered as crystal-like,
if the set of the BOO of this particle is close to the BOOs on the corresponding crystal. Otherwise we define the
particle as a liquid like.

We find that only in a pure A system spontaneous crystallization takes place ($53 \%$ of particles are
recognized as FCC-like and $25 \%$ as HCP-like. Others are liquid-like). In all other systems the number
of particles which are recognized as some of the crystals with symmetry found in genetic calculations
is at  a level of noise (at  least  $80 \%$ of the particles are liquid-like). Based on this observation
we conclude that spontaneous crystallization of the system into the structures obtained in genetic
algorithm calculations is strongly unprobable. For this reason it is difficult to find these structures
from simulations of liquids, which prevented the discovery of these structures before.

Figure 10 shows the number of nearest neighbors NN of the particles at $T=100$ K as a function of molar fraction of the B species. The partial number of
nearest neighbors of particles of different species is also shown. It is seen that NN slightly exceeds 12 which is the number of nearest neighbor of the
FCC crystal. Such situation is usually observed in a liquid at low temperature. Therefore this observation is also consistent with the absence of spontaneous
crystallization in the system.

Importantly, another set of crystal structures was observed in a study of the KA mixtures in Ref. \cite{Pedersen2018},
where two phase simulations of the KA mixture with different concentration of components were performed.
The disagreement of our results with the ones from \cite{Pedersen2018} can be caused by two reasons:
(i) the difference in the simulation setup (two phase simulations vs cooling of a single phase) or (ii)
metastable character of the structures reported in \cite{Pedersen2018}. As it is known, nucleation
of a liquid can result  in a metastable crystal which will change to a stable phase afterwards. However,
the transition from a metastable crystal to the stable one can require the time which is beyond the time of the
simulations in \cite{Pedersen2018}. Moreover, as it is seen from the convex hull construction (Fig. 1),
there are many structures with the energy close to the convex hull at $c_B>0.5$, therefore
strong competition between the different structures is expected, which should be the reason
of good glassforming ability of the KA system. For this reason the crystal structures obtained in genetic algorithm
calculations look more grounded.


\begin{table}
\begin{tabular}{ |c| c| c| c|c|c|c|}
\hline
 $c_B$ & Symmetry Group & $q_4$  & $q_6$ & $q_8$ & $q_{10}$  & $q_{12}$   \\
\hline
0.0 & 225 & 0.191 & 0.575 & 0.404 & 0.015 & 0.600 \\
\hline
 0.5 & 221 & 0.035 & 0.515 & 0.425 & 0.195 & 0.405 \\
\hline
 0.58 & 1 & 0.055  & 0.495 & 0.415 & 0.185 & 0.365  \\
\hline
 0.667 & 1 & 0.105 & 0.465 & 0.345 & 0.125 & 0.385 \\
\hline
 0.75 & 194 & 0.095 & 0.485 & 0.325 & 0.015 & 0.565 \\
\hline
1.0 & 194 & 0.097 & 0.485 & 0.317 & 0.010 & 0.565 \\
\hline
\end{tabular}
\caption{The values of different BOO in the crystal phases described in previous section. The first colomn gives the concentration
of the B specie of the structure, while the second one is the symmetry group. The results for $c_B=1.0$ correspond to the HCP structure, which is not the
equilibrium one, but a very close in the free energy.}
\end{table}

Figure 11 shows the shear and  the kinematic viscosity of the KA mixture at $T=100$ K and $P=425$ bar as functions of concentration $c_B$. Although the results are very noisy,
it is seen that both shear and kinematic viscosity are more or less smooth at the concentrations $c_B<0.5$ and very noisy at higher concentration of the B
component. This result is in good aggreement to the convex hull shown in Fig. 1, where we see that no stable crystal structures can be formed
at $c_B<0.5$ (except $c_B=0.0$), but several different crystals can be formed at higher $c_B$. Even if we do not observe the crystallization of the mixtures
in our simulations, based on the crystallization points of the pure components one can expect that the $T=100$ K isotherm should be close to the
freezing points of the system in the whole range of concentrations. The isothermal dependence of viscosity on the concentration usually demonstrates a
kink in the proximity of the point where the crystal structure is changed. For this reason we assume that noisy behavior of viscosity at $c_B>0.5$ is related
to the formation of the different crystalline structures in this range of concentrations.

\begin{figure}[htb]
\centering
\includegraphics[width=8cm]{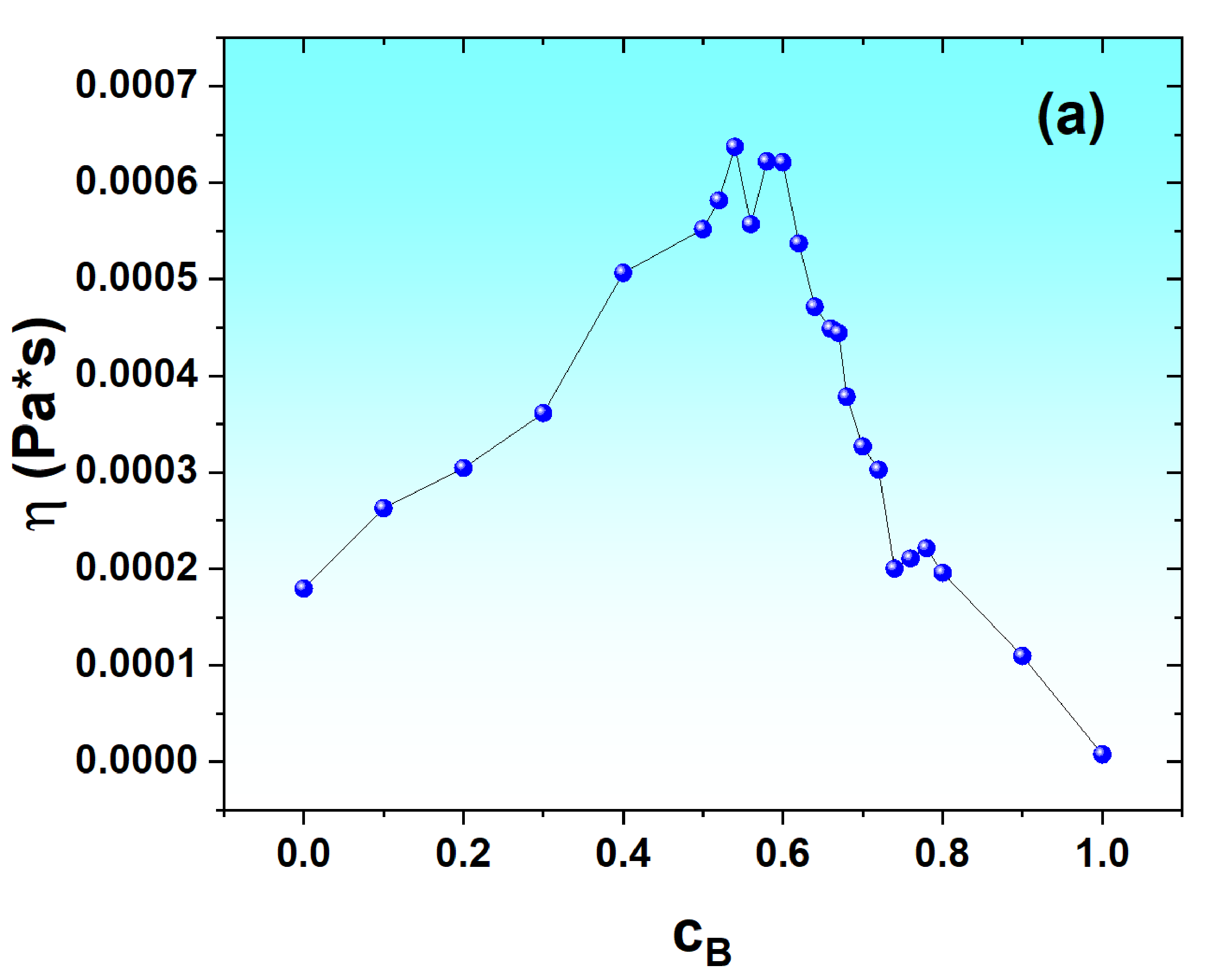}%

\includegraphics[width=8cm]{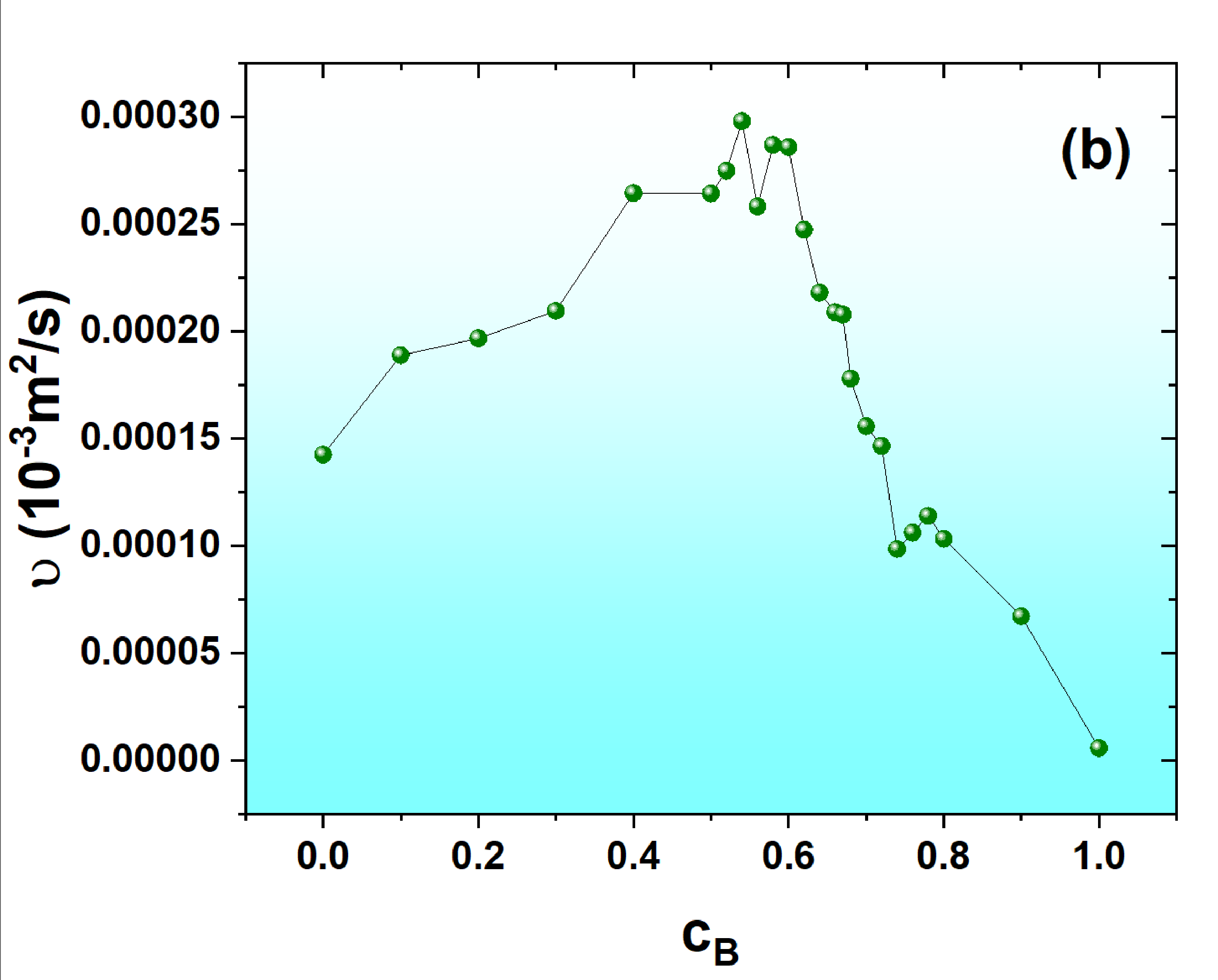}%

\caption{\label{visc} (a) Shear viscosity and (b) kinematic viscosity as a function of concentration of component B at $T=100$ K and $P=425$ bar. }
\end{figure}

\section{Conclusions}

In the present paper we perform a genetic algorithm search of the stable crystal phases in a binary LJ system with interactions of the Kob-Andersen mixture.
We find several stable crystal phases which correspond to the concentrations of the components $c_B=0$, $1/2$, $2/3$, $3/4$ and $1$.
At the same time these structures are not observed upon spontaneous crystallization of a liquid with given concentration. For this reason
these structures are difficult to obtain in a simple molecular dynamics simulation, which prevented their discovery before. At the same time
there are some indirect evidences of existence of the different crystal structures at $c_B>0.5$, for instance, the kinks on the isothermal dependence
of viscosity on the concentration of the components.

\end{document}